%% file: main.tex
\documentclass[twocolumn,times]{aastex631}

\usepackage{geometry}
\usepackage{enumitem}
\setlist{nosep}

\catcode`_=12 %
\renewcommand{\texttt}[1]{%
  \begingroup
  \ttfamily
  \begingroup\lccode`~=`-\lowercase{\endgroup\def~}{-\discretionary{}{}{}}%
  \begingroup\lccode`~=`_\lowercase{\endgroup\def~}{_\discretionary{}{}{}}%
  \catcode`-=\active\catcode`_=\active
  \scantokens{#1\noexpand}%
  \endgroup
}
\catcode`_=8 %

\usepackage{upquote}

\begin{document}
\title{SAGUARO: Time-domain Infrastructure for the Fourth Gravitational-wave Observing Run and Beyond}
\correspondingauthor{Griffin Hosseinzadeh}
\email{griffin0@arizona.edu}

\shorttitle{SAGUARO Forever!}
\shortauthors{Hosseinzadeh et al.}
\input{affiliation.tex}
\input{authors.tex}

\received{2023 October 12}
\revised{2024 January 16}
\accepted{2024 January 18}
\published{2024 March 13}
\submitjournal{The Astrophysical Journal}

\begin{abstract}
We present upgraded infrastructure for Searches after Gravitational waves Using ARizona Observatories (SAGUARO) during LIGO, Virgo, and KAGRA's fourth gravitational-wave (GW) observing run (O4).
These upgrades implement many of the lessons we learned after a comprehensive analysis of potential electromagnetic counterparts to the GWs discovered during the previous observing run.
We have developed a new web-based target and observation manager (TOM) that allows us to coordinate sky surveys, vet potential counterparts, and trigger follow-up observations from one centralized portal.
The TOM includes software that aggregates all publicly available information on the light curves and possible host galaxies of targets, allowing us to rule out potential contaminants like active galactic nuclei, variable stars, solar system objects, and preexisting supernovae, as well as to assess the viability of any plausible counterparts.
We have also upgraded our image-subtraction pipeline by assembling deeper reference images and training a new neural-network-based real--bogus classifier.
These infrastructure upgrades will aid coordination by enabling the prompt reporting of observations, discoveries, and analysis to the GW follow-up community, and put SAGUARO in an advantageous position to discover kilonovae in the remainder of O4 and beyond.
Many elements of our open-source software stack have broad utility beyond multimessenger astronomy, and will be particularly relevant in the ``big data'' era of transient discoveries by the Vera C.\ Rubin Observatory.
\end{abstract}



\section{Introduction}\label{sec:intro}
The discovery of a $\gamma$-ray burst (GRB) and a kilonova associated with gravitational waves (GWs) from the binary neutron-star (BNS) merger GW170817 ushered in a new era of multimessenger astronomy \citep{arcavi_optical_2017,coulter_swope_2017,ligo_scientific_collaboration_multi-messenger_2017,lipunov_master_2017,soares-santos_electromagnetic_2017,tanvir_emergence_2017,valenti_discovery_2017}. Among many other accomplishments, these observations confirmed that at least some GRBs come from neutron-star mergers \citep{goldstein_ordinary_2017,ligo_scientific_collaboration_gravitational_2017,savchenko_integral_2017}, revealed evidence of $r$-process nucleosynthesis in compact-object mergers \citep{chornock_electromagnetic_2017,mccully_rapid_2017,nicholl_electromagnetic_2017,pian_spectroscopic_2017,shappee_early_2017,smartt_kilonova_2017,tanvir_emergence_2017,watson_identification_2019}, and yielded the first measurement of the Hubble constant via the standard-siren method \citep{ligo_scientific_collaboration_gravitational-wave_2017}. Despite this breakthrough, little is known observationally about what diversity exists among kilonovae, caused either by differences in the progenitor system or by viewing-angle effects. The handful of kilonovae discovered in association with GRBs, and therefore viewed nearly pole-on, hints at an intrinsic luminosity span of up to two orders of magnitude \citep{Gompertz+18,Rastinejad+21,rastinejad_kilonova_2022,troja_nearby_2022,yang_long-duration_2022,yang_lanthanide-rich_2023,gillanders_heavy_2023,levan_jwst_2023}. A larger sample of off-axis kilonovae, discovered in association with GW events, is needed to fill in this parameter space.

\begin{figure*}
    \centering
    \includegraphics[width=0.706\textwidth]{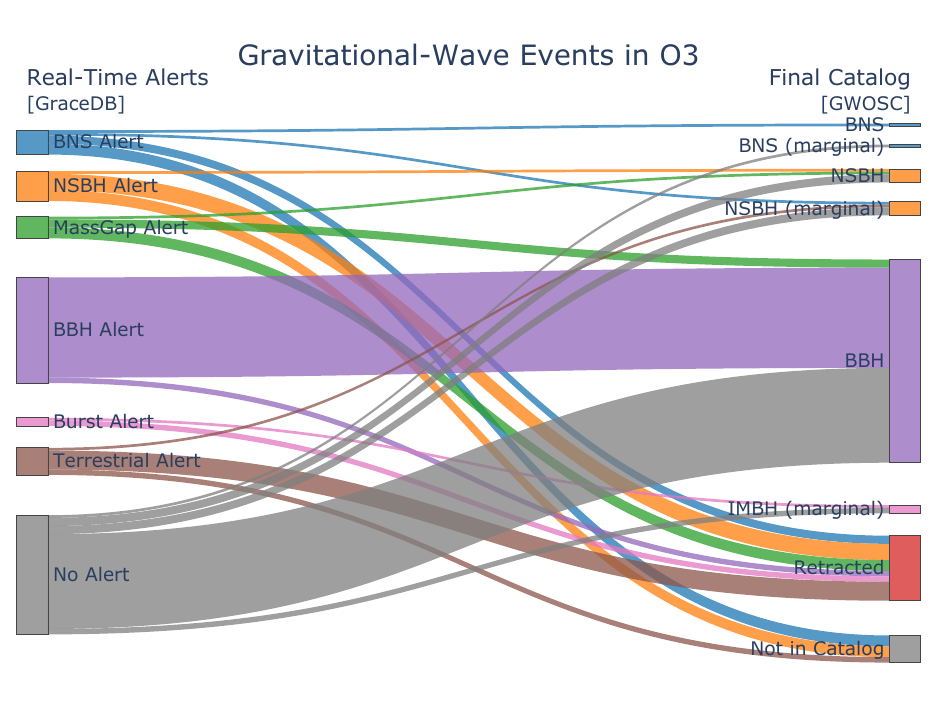}
    \caption{Sankey diagram showing the outcome of real-time alerts and the provenance of astrophysical GW events during LIGO and Virgo's third observing run, using data from the \href{https://gracedb.ligo.org/superevents/public/O3/}{Gravitational-Wave Candidate Event Database} and the \href{https://gwosc.org/eventapi/html/query/show?min-gps-time=1230336018&lastver=true}{Gravitational Wave Open Science Center}. Compact binary coalescence alerts are labeled by their highest probability source classification. Events with data-quality issues identified in real time are labeled as ``retracted.'' Events with a probability of astrophysical origin $p_\mathrm{astro} < 50\%$ after further analysis are labeled as ``marginal.'' Events with a false-alarm rate of $\mathrm{FAR} > 2\ \mathrm{yr}^{-1}$ after further analysis are not included in the final catalog \citep{GWTC-2,GWTC-3}. Only three confirmed and two marginal events containing an NS had alerts allowing follow-up in real time. SAGUARO covered a portion of the localization region for four of these events and discovered a single possible counterpart, which was unlikely to be a kilonova based on its luminosity \citep{paterson_searches_2021}. An interactive version of this figure that includes counts and event names for each flow is \href{https://content.cld.iop.org/journals/0004-637X/964/1/35/revision1/apjad2170f1_int/apjad2170f1_int.html}{available}.}
    \label{fig:sankey}
\end{figure*}

The Laser Interferometer GW Observatory (LIGO) and the Virgo GW detectors undertook their third observing run (O3) from 2019 April 1 to 2020 March 27, yielding detections of dozens of compact-object mergers, most of which were mergers of binary black holes \citep[BBH;][]{GWTC-2,GWTC-2.1,GWTC-3}. Figure~\ref{fig:sankey} summarizes the real-time alerts sent in O3 and a final assessment of their sources. Five events were promising targets for follow-up with electromagnetic (EM) telescopes, in that the lower-mass component of the binary had a mass ${<}3\ M_\sun$, and a detection alert was sent in nearly real time. Of those five, two ``confident'' and two ``marginal'' (referring to astrophysical significance) events were confirmed to be neutron-star--black-hole (NSBH) mergers: GW190426\_152155, GW190814, GW200105\_162426, and GW200115\_042309 \citep{ligo_scientific_collaboration_gw190814_2020,GWTC-2,GWTC-2.1,ligo_scientific_collaboration_observation_2021,GWTC-3}. The fifth (GW190425) was confirmed as a confident BNS merger \citep{abbott_gw190425_2020}. Despite copious follow-up, no EM transients were confirmed as GW counterparts (\citealt{Andreoni+19,Andreoni+20,Coughlin+19,Dobie+19,dobie_comprehensive_2022,Goldstein+19,Gomez+19,Hosseinzadeh+19,lundquist_searches_2019,Ackley+20,antier_first_2020,Antier+20b,Garcia+20,Gompertz+20b,Kasliwal+20,Morgan+20,Pozanenko+20,Thakur+20,Vieira+20,Watson+20,Alexander+21,Anand+21,Becerra+21,Bhakta+21,Chang+21,de_Wet+21,Dichiara+21,Kilpatrick+21,Oates+21,Ohgami+21,paterson_searches_2021,deJaeger+22,tucker_soar_2022}; see \citealt{rastinejad_systematic_2022} for a comprehensive discussion).

Our EM follow-up program, Searches After Gravitational waves Using ARizona Observatories (SAGUARO), surveys the localization regions of GW events by partnering with the Catalina Sky Survey (CSS), which routinely searches for near-Earth asteroids \citep{christensen_catalina_2019}.
CSS's 1.5~m telescope on Mt.~Lemmon, Arizona, surveys fixed 5~deg$^2$ fields with a typical unfiltered limiting magnitude of $\sim$21.5~mag.
In the event of a GW alert requiring urgent follow-up, we have the ability to send a ranked list of fields to the CSS's scheduling software, which will interrupt normal survey operations on the 1.5~m to observe the GW localization as soon as possible. Serendipitous CSS observations may also cover parts of GW localization regions without our request.

If we or others discover a plausible EM counterpart, we will trigger targeted follow-up observations under our approved program on any or all of the following observatories:
\begin{enumerate}
    \item the $2 \times 8.4$~m Large Binocular Telescope on Mt.~Graham, Arizona, USA,
    \item the 6.5~m Magellan Telescopes at Las Campanas Observatory, Chile,
    \item the 6.5~m MMT on Mt.~Hopkins, Arizona, USA,
    \item the 2.3~m Bok Telescope on Kitt Peak, Arizona, USA,
    \item the 1.8~m Vatican Advanced Technology Telescope on Mt.~Graham, Arizona, USA, and
    \item the 1.5~m Kuiper Telescope on Mt.~Bigelow, Arizona, USA.
\end{enumerate}
We also participate in active GW follow-up programs on several facilities worldwide, which protects our effort against bad weather at a single site and provides us the ability to obtain follow-up photometry and spectroscopy of potential kilonovae within minutes to hours of discovery.

\cite{lundquist_searches_2019} present an overview of the SAGUARO's infrastructure as it stood at the beginning of O3, and \cite{paterson_searches_2021} present an analysis of the O3 follow-up effort, including 17 GW events with CSS data either triggered by our program or observed serendipitously as part of the normal near-Earth asteroid survey.
We continue this series of papers by describing new and improved hardware and software infrastructure to be used by SAGUARO during LIGO, Virgo, and KAGRA's fourth observing run (O4), which started on 2023 May 24. In Section~\ref{sec:tom}, we present our new target and observation manager (TOM), an interactive web front end and database back end for carrying out the project. In Section~\ref{sec:pipeline}, we describe updates to our image-subtraction pipeline and machine-learning algorithm for real--bogus classification. In Section~\ref{sec:future}, we outline our future plans, including progress toward adding a new instrument to our program. In Section~\ref{sec:summary}, we conclude by discussing our prospects for discovering the second kilonova during O4. All of our software, including installation instructions, is publicly available on GitHub\footnote{\url{https://github.com/SAGUARO-MMA/}} and archived on Zenodo \citep{daly_sassy_2023,hosseinzadeh_saguaro_2023,hosseinzadeh_tom_2023,paterson_saguaro_2023,rastinejad_kilonova_2023}.

\section{Target and Observation Manager}
\label{sec:tom}
The complexity of our observing program and the necessity of responding in real time to alerts and incoming data require that we centralize all of our tools for planning observations, monitoring data, vetting targets, and reporting results in a single location that is accessible to all team members at all times. Current transient surveys and large follow-up programs use similar systems, including the Asteroid Terrestrial-impact Last Alert System (ATLAS) Transient Science Server \citep{smith_design_2020}, the Global Rapid Advanced Network Devoted to the Multi-messenger Addicts' Interface and Communication for Addicts of the Rapid follow-up in multi-messenger Era \citep{antier_first_2020}, the Global Relay of Observatories Watching Transients Happen Marshal \citep{kasliwal_growth_2019}, the Global Supernova Project's Supernova Exchange \citep{howell_global_2017}, the Public ESO Spectroscopic Survey of Transient Objects Marshall \citep{smartt_pessto:_2015}, the Young Supernova Experiment's YSE-PZ \citep{coulter_yse-pz_2023}, and the Zwicky Transient Facility's Fritz \citep[based on SkyPortal;][]{coughlin_data_2023}.

We call our portal a target and observation manager (TOM), after \cite{street_general-purpose_2018}. The SAGUARO TOM is a browser-based front end on top of a PostgreSQL~14 database,\footnote{\url{https://www.postgresql.org/}} built using the Django\footnote{\url{https://www.djangoproject.com/}}-based TOM Toolkit \citep{lindstrom_tom_2022} and its affiliated packages. The TOM is hosted on a dedicated server located on the University of Arizona campus and is available over the web to all members of our team with login credentials. In the following sections, we describe the various aspects of our observing campaign that have now been integrated into the TOM.

\subsection{Receiving GW Alerts}\label{sec:alerts}
\begin{figure*}
    \centering
    \includegraphics[width=\textwidth,trim={0 2.3in 0 0},clip]{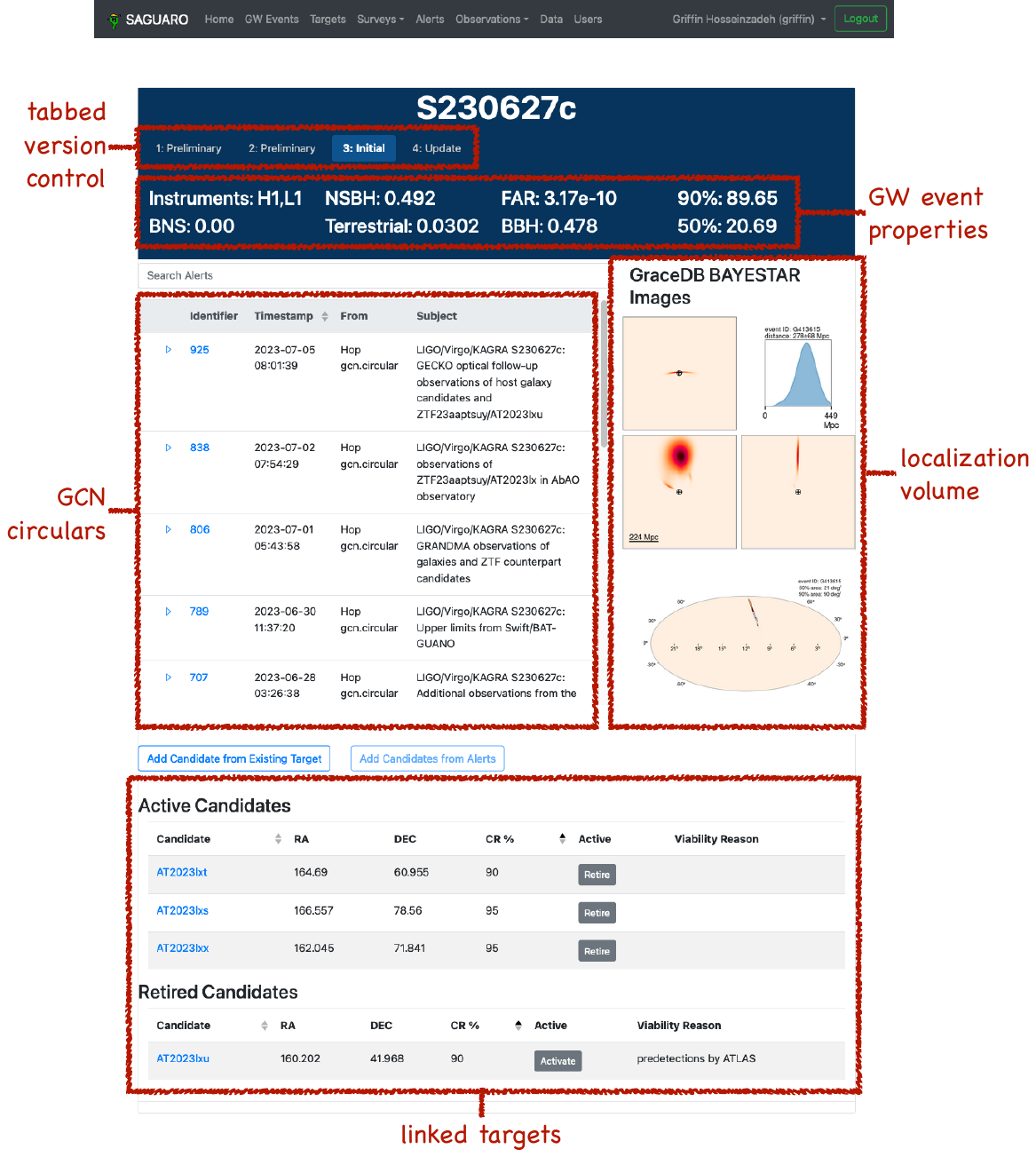}
    \caption{GW event detail page from the SAGUARO TOM, showing the source classification probabilities, false-alarm rate, localization areas, and GCN circulars for event S230627c. The tabbed interface switches between different versions of the alert information. The active and retired candidates lists at the bottom of the page link to targets that have been suggested as low-probability EM counterparts to the GW event by \cite{anumarlapudi_ligo_2023}.}
    \label{fig:nonlocalizedevent_detail}
\end{figure*}

Our workflow begins with the receipt of an alert from the LIGO/Virgo/KAGRA Collaboration notifying us of a GW event. Starting in O4, the International Gravitational Wave Observatory Network\footnote{\url{https://emfollow.docs.ligo.org/userguide/}} has been sending alerts over Hopskotch,\footnote{\url{https://scimma.org/hopskotch}} an Apache Kafka\footnote{\url{https://kafka.apache.org/}}-based alert platform developed by the Scalable Cyberinfrastructure to support Multi-Messenger Astrophysics (SCIMMA) project.\footnote{\url{https://scimma.org/}} Two packages affiliated with the TOM Toolkit allow us to receive and handle these alerts. The \texttt{tom-alertstreams} app\footnote{\url{https://github.com/TOMToolkit/tom-alertstreams}} provides a management command \verb|readstreams| that listens for Hopskotch alerts and passes the alert contents to a user-defined handler function. The \texttt{tom_nonlocalizedevents} app\footnote{\url{https://github.com/TOMToolkit/tom_nonlocalizedevents}} provides basic handler functions that store alert information in the TOM database and web templates to display information for a given GW event, including the contents of related circulars from the General Coordinates Network (GCN)\footnote{\url{https://gcn.nasa.gov/}} retrieved via SCIMMA's Hermes message exchange service\footnote{\url{https://hermes.lco.global/?topic=gcn.circular}} (Figure~\ref{fig:nonlocalizedevent_detail}; see also Section~\ref{sec:hermes}).

\begin{figure}
    \centering
    \includegraphics[width=\columnwidth]{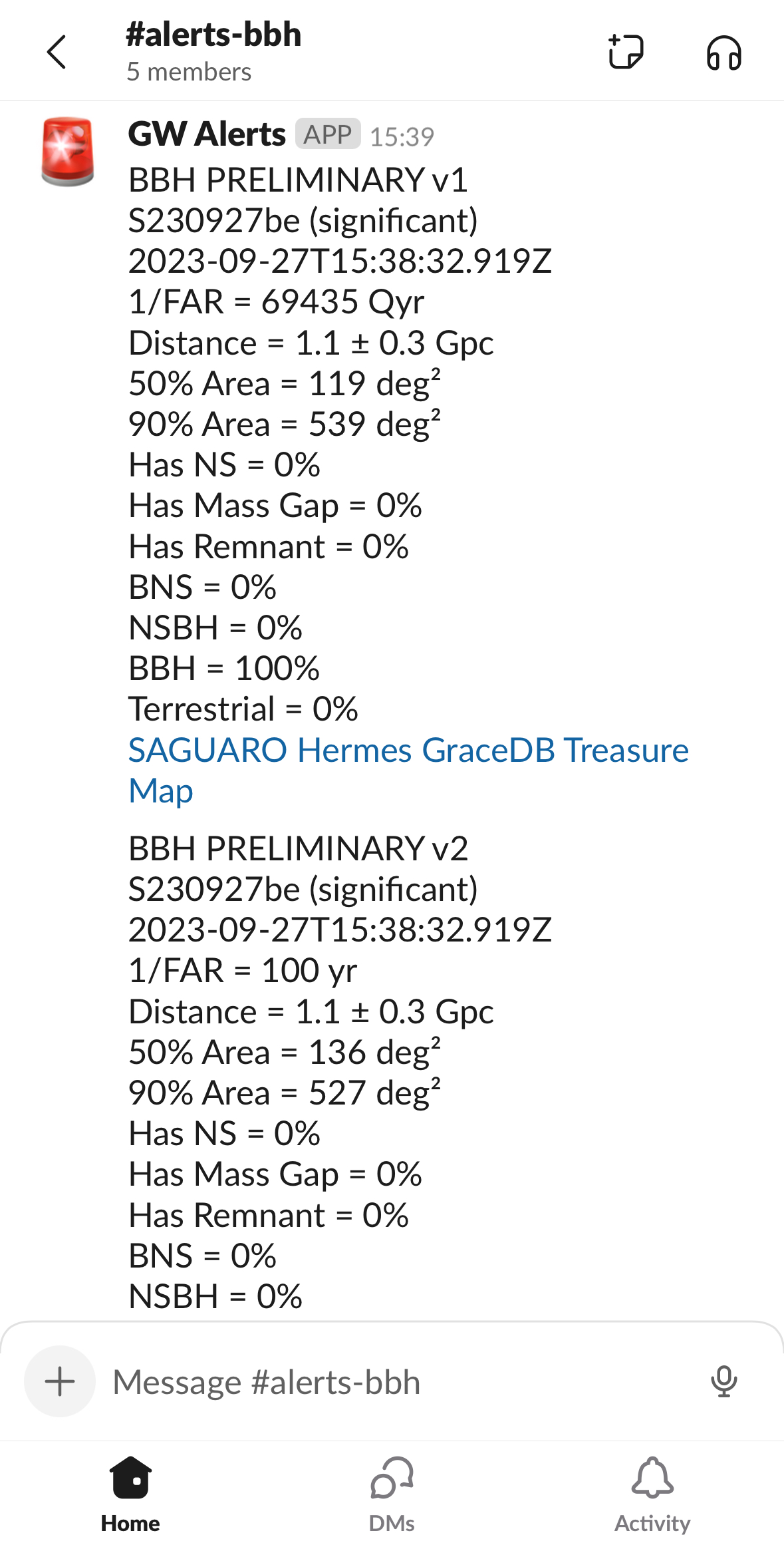}
    \caption{Real-time human-readable alert summary sent by the SAGUARO TOM, as viewed on the Slack mobile app. The message time stamp shows that the first alert arrived within $\sim$1 minute of the detection. The summary contains helpful information for determining whether follow-up is appropriate, including the inverse of the false-alarm rate (1/FAR), the distance to the source, the areas of the 50\% and 90\% localization regions, and, for compact binary coalescences, the probabilities of the progenitor system containing at least one object with mass ${<}3\ M_\sun$ (Has NS) or $3{-}5\ M_\sun$ (Has Mass Gap), the event having an EM counterpart (Has Remnant), and the signal coming from a BNS merger, NSBH merger, BBH merger, or terrestrial noise source. More information is easily accessed by following the hyperlinks at the bottom of the summary.}
    \label{fig:slack}
\end{figure}

We customized the alert handler functions in our TOM to provide additional functionality. We have the ability to send human-readable summaries of each event over short message service (SMS) through Twilio\footnote{\url{https://www.twilio.com/}} and Slack through a custom app\footnote{\url{https://api.slack.com/start/apps}} (Figure~\ref{fig:slack}). We divide incoming alerts into five streams, with the conditions in parentheses evaluated in the following order:
\begin{enumerate}
    \item test alerts (if the event name starts with ``MS''), 
    \item subthreshold alerts (if \verb|significant=False|),\label{item:subthreshold}
    \item burst alerts (if \verb|group='Burst'|),
    \item BBH alerts (if $\mathtt{HasNS} < 1\%$ and $\mathtt{NSBH} < 1\%$ and $\mathtt{BNS} < 1\%$), and 
    \item NS alerts (otherwise).\label{item:ns}
\end{enumerate}
Each user can choose which SMS streams to subscribe to via their profile page in the TOM, and streams \ref{item:subthreshold}--\ref{item:ns} each have dedicated channels in our Slack workspace. NS alerts push notifications to all members of the channel with the \verb|@channel| command.

The new Kafka alerts contain multiorder coverage \citep[MOC;][]{fernique_moc_2014} maps of the localization area in the Hierarchical Equal Area isoLatitude Pixelization (HEALPix) format \citep{gorski_healpix_2005}. The \texttt{tom_nonlocalizedevents} package stores these maps in the TOM database with the help of HEALPix Alchemy \citep{singer_healpix_2022}. For alerts containing a new sky map, including maps generated by external coincidence with a GRB, we calculate and store the innermost credible region (among 25\%, 50\%, 75\%, 90\%, and 95\%) that contains the coordinates of each linked target (see Section~\ref{sec:vetting}) and the center of each survey field. For computational efficiency, we use the center of the field as a proxy for all future sources we discover in that field. We also calculate and store the 90\% credible region contour itself, which we use for display (see Section~\ref{sec:survey}). Last, we calculate and store the probability contained within the footprints of each of our survey fields, which we use to prioritize our observation requests. This last calculation requires rasterization of the MOC map using \texttt{healpy} \citep{zonca_healpy_2019} and \texttt{ligo.skymap} \citep{singer_rapid_2016}.

\subsection{Planning and Triggering a Search}\label{sec:survey}
\begin{figure*}
    \centering
    \includegraphics[width=\textwidth,trim={0 4.3in 0 0},clip]{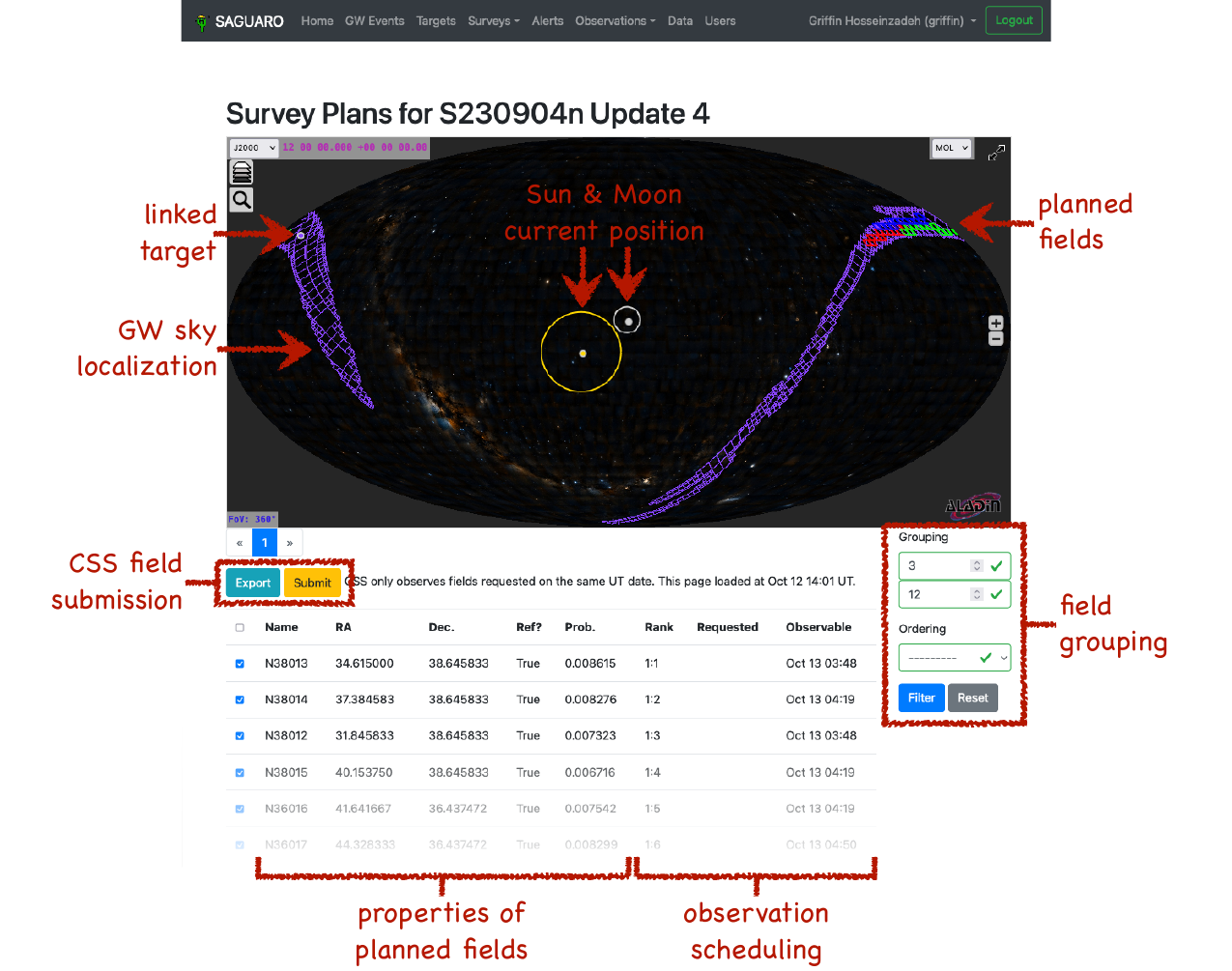}
    \caption{Survey planning page on the SAGUARO TOM, including an Aladin Lite viewer showing three groups of 12 planned fields (red, green, and blue), the 90\% credible region contour for S230904n (purple), and the current positions of the sun (gold) and moon (silver). The radii of the sun and moon markers roughly represent their respective exclusion regions, although these are distorted with respect to the Mollweide projection. The purple dot near the top left of the map shows AT~2023rkw, suggested as a possible EM counterpart by \cite{necker_ligo_2023} but then classified as a Type~Ia supernova by \cite{anand_ligo_2023}. The planned fields are listed in a table at the bottom of the page, with their probability, rank order in our plan, time requested (if any), and time first observable in UT.}
    \label{fig:survey}
\end{figure*}

Once we decide to follow up a GW event, we use the TOM to plan our survey and send instructions to the CSS 1.5~m telescope. CSS observes in groups of 12 adjacent fields as part of their near-Earth asteroid survey, so we generate survey plans using the following algorithm: starting with the highest probability field in the localization region, we select the next highest probability adjacent field that is observable between now and the next sunrise (using 12\degr{} twilight), taking into account an airmass limit of 1.75 (to allow time for the full 12-field pattern to complete) and a phase-dependent moon exclusion region of $(3 + 42\phi)\degr$ while the moon is above the horizon, where $\phi \in [0, 1]$ is the moon phase. We use the Astropy \texttt{coordinates} package \citep{astropy_collaboration_astropy_2022} for visibility calculations and Astroplan \citep{morris_astroplan_2018} to look up the moon phase. We iterate, excluding fields for which we do not have reference images (see Section~\ref{sec:references}), until we reach 12 fields (60 deg$^2$). Depending on the size and position of the localization region, we may generate and submit multiple sets of 12 fields. To visualize our survey plans, we embed version 3 of the Aladin Lite sky viewer \citep{baumann_aladin_2022}, which natively reads HEALPix maps, into our TOM's survey planning page. An example is shown in Figure~\ref{fig:survey}.

Survey plans can be manually edited by selecting or deselecting fields using checkboxes in the table. Users can then click the yellow ``Submit'' button on the survey planning page to submit the list of fields to CSS. This button opens one or more files on CSS's server using Paramiko\footnote{\url{https://www.paramiko.org/}} and writes lists of 12 field names. These fields are then ingested by CSS's scheduling software and observed as soon as they are visible. Users can also click the teal ``Export'' button to download a copy of the field lists to their local computer.

As we have plans to expand to other telescopes (see Section~\ref{sec:future}), we have taken care to make our survey planning infrastructure as general as possible. Additional survey fields can be added for different facilities, including fields centered on galaxies for imagers with small fields of view. We plan to release this module as a \texttt{tom_surveys} app for the TOM Toolkit. The core of this app consists of the \verb|SurveyField| and \verb|SurveyObservationRecord| models, which are analogous to the \verb|Target| and \verb|ObservationRecord| models included with the base TOM Toolkit but without the connection to a single pair of coordinates implied by a ``target.''

\subsection{Scanning Survey Candidates}\label{sec:scanning}
We copy all CSS images to our data-reduction server at 5~minute intervals, regardless of whether we requested them.
In some cases, their ongoing near-Earth asteroid survey will overlap with GW localization regions by coincidence; we refer to these as ``serendipitous'' observations.
\cite{paterson_searches_2021} describe our data-reduction pipeline in detail; it performs image subtraction and detects new sources in the difference images, which we refer to as ``survey candidates'' because, at this stage, they may either be real transients or artifacts from the subtraction. We use the optimal image-subtraction algorithm of \citet{zackay_proper_2016}, which takes in a new image and a deep reference image of the same field and produces a raw difference image and a significance image corrected for astrometric noise (called $S_\mathrm{corr}$). We refer to this set of four images as ``image quadruplets'' below. Updates to the reference images and real--bogus machine-learning classifier are discussed in Section~\ref{sec:pipeline}.

\begin{figure*}
    \centering
    \includegraphics[width=\textwidth,trim={0 2.9in 0 0},clip]{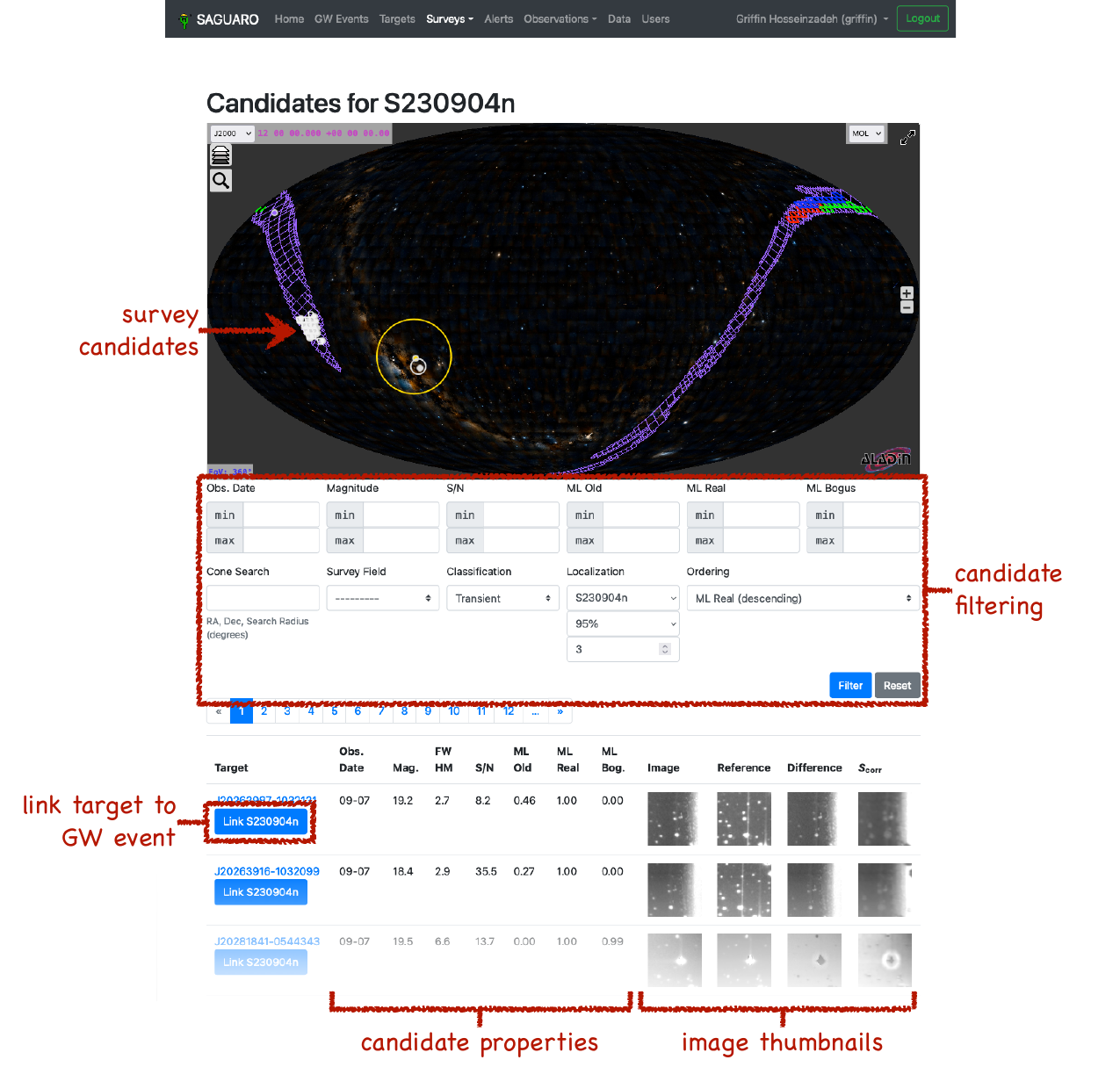}
    \caption{Survey candidates page for S230904n in the SAGUARO TOM. As on the survey planning page in Figure~\ref{fig:survey}, the Aladin sky map shows the 90\% localization region (purple) and current sun (gold) and moon (silver) position. Observed fields and detections are shown in white.
    Below, the survey candidates table shows a temporary coordinate-based name, the observation month and day, magnitude, full width at half maximum (FWHM), signal-to-noise ratio (S/N), three real--bogus machine-learning (ML) scores, and thumbnails of the image quadruplet. The blue ``Link'' buttons allow the user to bookmark promising sources for further vetting in association with the chosen GW event.}
    \label{fig:candidates}
\end{figure*}

Once detections are in the database, they can be viewed, filtered, and sorted in the TOM. The survey candidates page (Figure~\ref{fig:candidates}) shows the observed fields and candidate point sources overlaid on the localization region in the Aladin sky map for a chosen GW event, followed by a table of sources with their real--bogus scores (see Section~\ref{sec:ml}) and image quadruplets (see Section~\ref{sec:ingestion}). By default, we display sources within the 95\% localization region of the event observed between 0 and 3 days after the merger time, sorted by decreasing ``real'' score, although further filtering and sorting is available via the form above the table. It is also possible to remove the filtering by localization region and view all detections from the previous night, as would be done for a typical all-sky time-domain survey. If a detection looks real, the user can link it to the GW event by clicking the blue ``Link'' button below its temporary target name, effectively bookmarking it for further vetting.

\subsection{Vetting Potential Kilonovae}
\label{sec:vetting}

Motivated by the large number of contaminating sources found during searches following O3 merger events, we incorporate a number of publicly available tools into our infrastructure that automatically vet potential kilonovae, from both our own survey and others. In \cite{rastinejad_systematic_2022}, we evaluated the performance of each tool on potential O3 kilonovae and made recommendations for including automatic vetting in O4 follow-up. Here, we describe the components of our Kilonova Candidate Vetting package,\footnote{\url{https://github.com/SAGUARO-MMA/kne-cand-vetting}} which implements the procedures described in our previous work \citep{rastinejad_systematic_2022} for SAGUARO's O4 follow-up effort.

We have downloaded and ingested each of the catalogs we use for vetting into the TOM database, so that we can quickly run queries for large numbers of targets, and indexed each catalog table using the Quad Tree Cube \citep[Q3C;][]{koposov_q3c_2006} PostgreSQL extension for efficient cone searches. The entire procedure below, with the exception of ATLAS forced photometry, has been wrapped into a single function, which is run automatically in two situations: when a button is pressed on the target detail page, useful for potential kilonovae from our own survey that already exist in the TOM, or when a new target is added through the TOM web interface, useful for potential kilonovae reported by other surveys.  Because ATLAS forced photometry can take several minutes, we run it asynchronously on the press of a separate button.

\begin{figure*}
    \centering
    \includegraphics[width=\textwidth,trim={0 2in 0 0},clip]{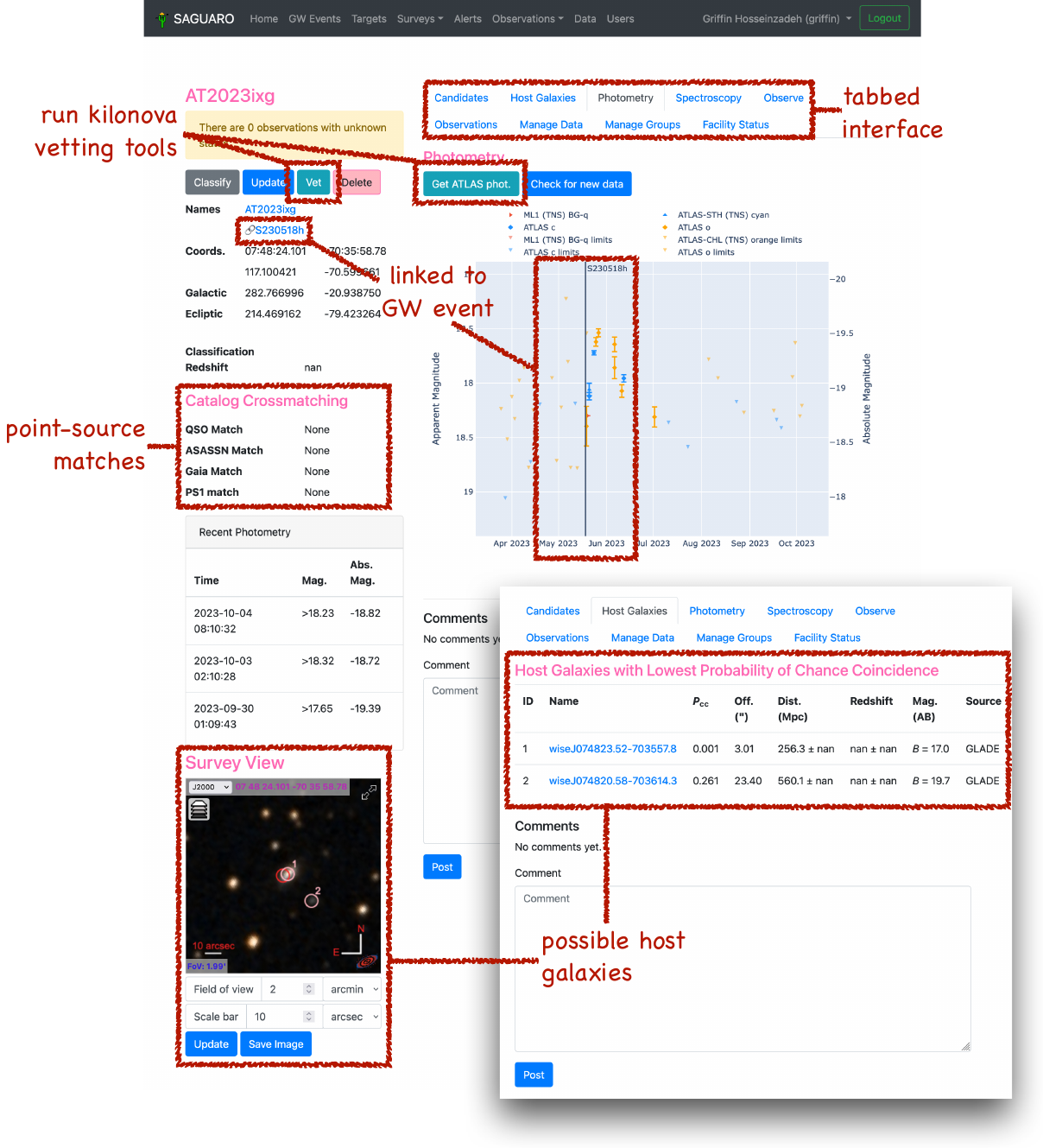}
    \caption{Two tabs of the target detail page for AT~2023ixg, which was suggested as a possible EM counterpart to GW event S230518h by \cite{johnston_two_2023}. The teal ``Vet'' button near the top left of the page runs our kilonova vetting code. The results are displayed in the ``Catalog Crossmatching'' section at the center left of the page, which indicates no point-source matches but several galaxy matches in this case. The ``Photometry'' tab shows a plot of all known photometric detections and upper limits for this target, including public data from the ATLAS forced photometry server (obtained by pressing the teal ``Get ATLAS phot.''\ button), the Transient Name Server, and the Zwicky Transient Facility. Because this target has been linked with GW candidate S230518h (note the chain icon under the target names), the time of merger of that event is indicated by a vertical line on the photometry plot. The ``Host Galaxies'' tab (inset) shows a table of potential host galaxies, ordered by increasing probability of chance coincidence ($P_\mathrm{cc}$). These are also marked on the Aladin viewer at the bottom left of the page.}
    \label{fig:target}
\end{figure*}

\subsubsection{Point-source Matching}

We begin by crossmatching each target with several point-source catalogs to identify known variable sources (namely, quasars, active nuclei, and variable stars) that may masquerade as kilonovae. We perform a cone search on each catalog using a $2''$ radius, chosen to match the typical seeing of ground-based surveys, and then apply additional cuts based on the specification of each catalog. Where a match is determined, we display the angular offset, catalog probability score (where available), and source class (e.g., variable star type, where available) on the target detail page (see Figure~\ref{fig:target}). This page also includes a view of the field around the target via an embedded Aladin Lite viewer \citep[version 2;][]{boch_aladin_2014}, which we can use to identify nearby point sources or false matches by eye.

To identify quasar matches, we utilize the Million Quasars Catalog \citep[Milliquas;][]{flesch_million_2021} compilation, which includes quasars from SDSS DR16 \citep{ahumada_16th_2020}. Following \citet{Flesch15} we rule out targets as potential kilonovae if they have a $>$97\% probability of being a quasar. 
To search for variable star contaminants, we employ three unique catalogs. First, we look for matches within the Gaia Data Release 3 variable star catalog (DR3; \citealt{{Gaia_DR3_2023,Gaia_DR3_VariableStar}}; updated from using eDR3 in \citealt{rastinejad_systematic_2022}). In addition, we crossmatch with the All-Sky Automated Survey for Supernovae \citep[ASAS-SN;][]{jayasinghe_asassn_2019} variable star catalog. We rule out all targets as kilonovae that have matches in either catalog.
Finally, we search the Pan-STARRS1 (PS1) $3\pi$ Survey \citep{chambers_pan-starrs1_2016} point-source catalog \citep{tachibana_morphological_2018}. The catalog provides a point-source probability score for nearly all sources in PS1 using a random-forest machine-learning (ML) algorithm \citep{breiman_random_2001}. Following \citet{tachibana_morphological_2018}, we utilize an ML score cutoff of 0.83 to determine if matches are point sources and thus unrelated to the GW merger event.

\subsubsection{Host-galaxy Matching}

We incorporate five individual galaxy catalogs to further contextualize each target and determine if its distance and luminosity are consistent with expectations of a kilonova counterpart. In \citet{rastinejad_systematic_2022} we demonstrated that host-galaxy information was powerful in determining if targets were associated with GW events, as it ruled out $>$30\% of potential O3 kilonovae without the need for any follow-up observations. We begin by querying each catalog for a match within $3\farcm44$ of the target. This radius translates to a 100~kpc offset at a distance of 100~Mpc, consistent with the upper end of short GRB offsets (though the median offset lies $\sim$8~kpc; e.g., \citealt{Fong+22}). To determine the likelihood the target is associated with each source, we utilize the probability of chance coincidence method \citep[$P_{\rm cc}$;][]{Bloom+02}. $P_{\rm cc}$ combines the galaxy's offset and magnitude, and is used to associate GRBs (including NS merger-origin GRBs) to host galaxies (e.g., \citealt{Bloom+02}). Where available, we calculate $P_{\rm cc}$ using the galaxy's $r$-band magnitude, as it is generally the most widely available filter, though in cases where this filter is not provided we employ a band with similar wavelength coverage. We display the 10 galaxies with the lowest $P_{\rm cc}$ values and $P_{\rm cc} < 0.8$ on each target's page and circle them in the Aladin viewer. For each potential host galaxy, we display its name, $P_{\rm cc}$, offset from the target, distance or redshift (depending on what the catalog provides), magnitude, and catalog source (see Figure~\ref{fig:target}, bottom right).

As we expect counterparts to an O4 GW event to originate in a low-redshift ($z \lesssim 0.05$) galaxy, we employ three galaxy catalogs specifically designed for the local Universe. Using the method described above, we query the Gravitational Wave Galaxy Catalog \citep[GWGC;][]{white_list_2011}, the Galaxy List for the Advanced Detector Era \citep[GLADE+;][]{dalya_glade_2018,dalya_glade_2022}, and the Heraklion Extragalactic Catalog \citep[HECATE;][]{kovlakas_heraklion_2021} for galaxies with known distances. Using $B$-band luminosity density, the GLADE+ catalog is complete out to $d_L=47^{+4}_{-2}$~Mpc \citep{dalya_glade_2022} while HECATE is $>$50\% complete up to $d_L = 170$~Mpc \citep{kovlakas_heraklion_2021}. In addition, to identify contaminants that may be associated with higher-redshift galaxies, we use photometric redshifts from the Sloan Digital Sky Survey Data Release 12 \citep[SDSS DR12;][]{alam_eleventh_2015} and the Pan-STARRS Source Types and Redshifts with Machine Learning \citep[PS1-STRM;][]{Beck+21} catalog. SDSS DR12 also provides a number of spectroscopic redshifts that we incorporate when available.

\subsubsection{Survey Photometry Matching}

Finally, we search for and display any additional survey photometry for each target. This survey data is useful for determining if the transient was detected prior to the merger (ruling it out as a counterpart) and provides information on the color and temporal evolution. We again employ a search radius of $2''$ to match the typical seeing at ZTF's Palomar Observatory. We automatically ingest photometry from the public ZTF alert stream \citep{bellm_zwicky_2019,masci_zwicky_2019,patterson_zwicky_2019} and from the Transient Name Server \citep[TNS;][]{gal-yam_tns_2021}. We also gather ATLAS data using their forced photometry service \citep{tonry_atlas_2018,smith_design_2020,shingles_release_2021}, and stack and sigma-clip observations from the same night using the script they provide.\footnote{\url{https://gist.github.com/thespacedoctor/86777fa5a9567b7939e8d84fd8cf6a76}} Each of these streams provides photometry from difference images.

\subsection{Coordinating Follow-up}\label{sec:follow-up}
Once the target has been vetted, rapid photometric and spectroscopic follow-up is crucial to confirm the association of the EM counterpart and gather early data for later analysis of the kilonova. The TOM Toolkit already includes the infrastructure to submit observations to all facilities in the Astrophysical Events Observatories Network \citep[AEON;][]{street_astrophysical_2020}. Currently, this includes optical and near-infrared imagers and spectrographs on Las Cumbres Observatory \citep{brown_cumbres_2013}, SOAR \citep{clemens_goodman_2004,schlawin_design_2014}, and Gemini \citep{hook_gemini-north_2004,eikenberry_flamingos-2_2006,elias_design_2006,elias_performance_2006}, all of which have active GW follow-up programs in which we participate (see Section~\ref{sec:intro}).

\begin{figure*}
    \centering
    \includegraphics[width=0.706\textwidth]{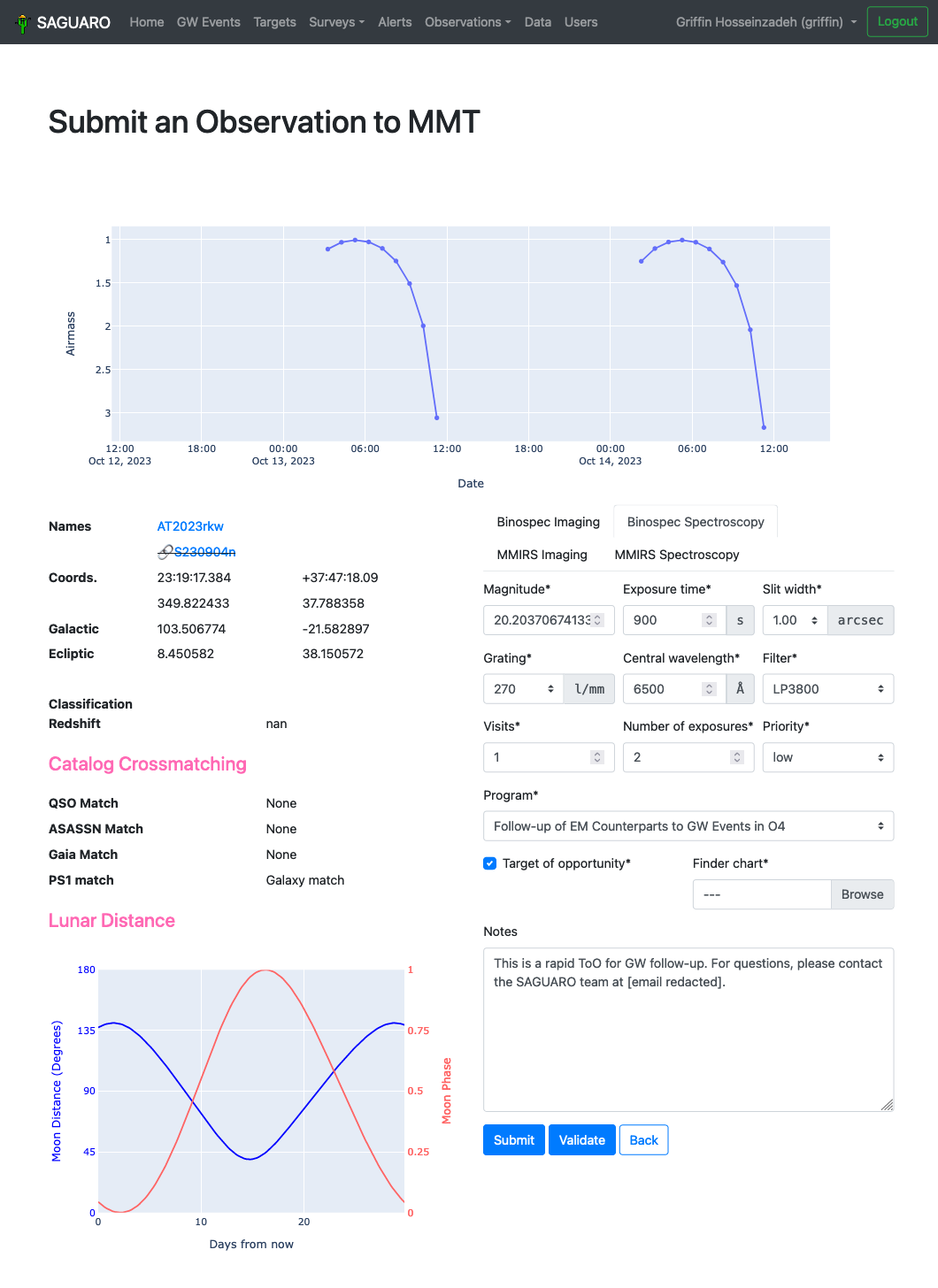}
    \caption{MMT observation request page for AT~2023rkw on the SAGUARO TOM, provided by our \texttt{tom_mmt} module. AT~2023rkw was suggested as a possible EM counterpart to S230904n by \cite{necker_ligo_2023} but then classified as a Type~Ia supernova by \cite{anand_ligo_2023}. Therefore, the link to the GW event at the left of the page is struck through. An airmass chart (top) and lunar distance chart (bottom left) show the visibility of the target. The form at the bottom right of the page allows the user to specify the instrument settings, choose a program, upload a finder chart, and provide free-text notes. The tabbed interface provides four forms, for imaging and spectroscopy on Binospec and MMIRS.}
    \label{fig:mmt}
\end{figure*}

In our TOM, we have added the ability to submit observations to two imaging spectrographs on the MMT: Binospec in the optical \citep{fabricant_binospec_2019} and the MMT and Magellan Infrared Spectrograph in the near-infrared \citep[MMIRS;][]{mcleod_mmt_2012}. The \texttt{tom_mmt} app\footnote{\url{https://github.com/griffin-h/tom_mmt}} is an ``observatory module'' for the TOM Toolkit based on PyMMT \citep{wyatt_pymmt_2023,shrestha_evidence_2024}, a recently released Python package that communicates with the application programming interface (API) of the MMT Observatory Manager \citep{gibson_queue_2018}. The \texttt{tom_mmt} module provides observation request forms (Figure~\ref{fig:mmt}, bottom), which are prefilled with default settings based on observations of AT~2017gfo \citep{coulter_swope_2017} and theoretical models of kilonovae \citep[e.g.,][]{kasen_origin_2017}, along with the most recent magnitude of the target in question. The module also adds the MMT to the TOM's airmass charts (Figure~\ref{fig:mmt}, top) and ``Facility Status'' page, allowing the user to quickly determine which instrument is available for the upcoming night. Last, it allows the user to check the status of requested observations without leaving the TOM.

\subsection{Visualizing Data}\label{sec:data}
\begin{figure}
    \centering
    \includegraphics[width=\columnwidth,trim={0 0.5in 0 0},clip]{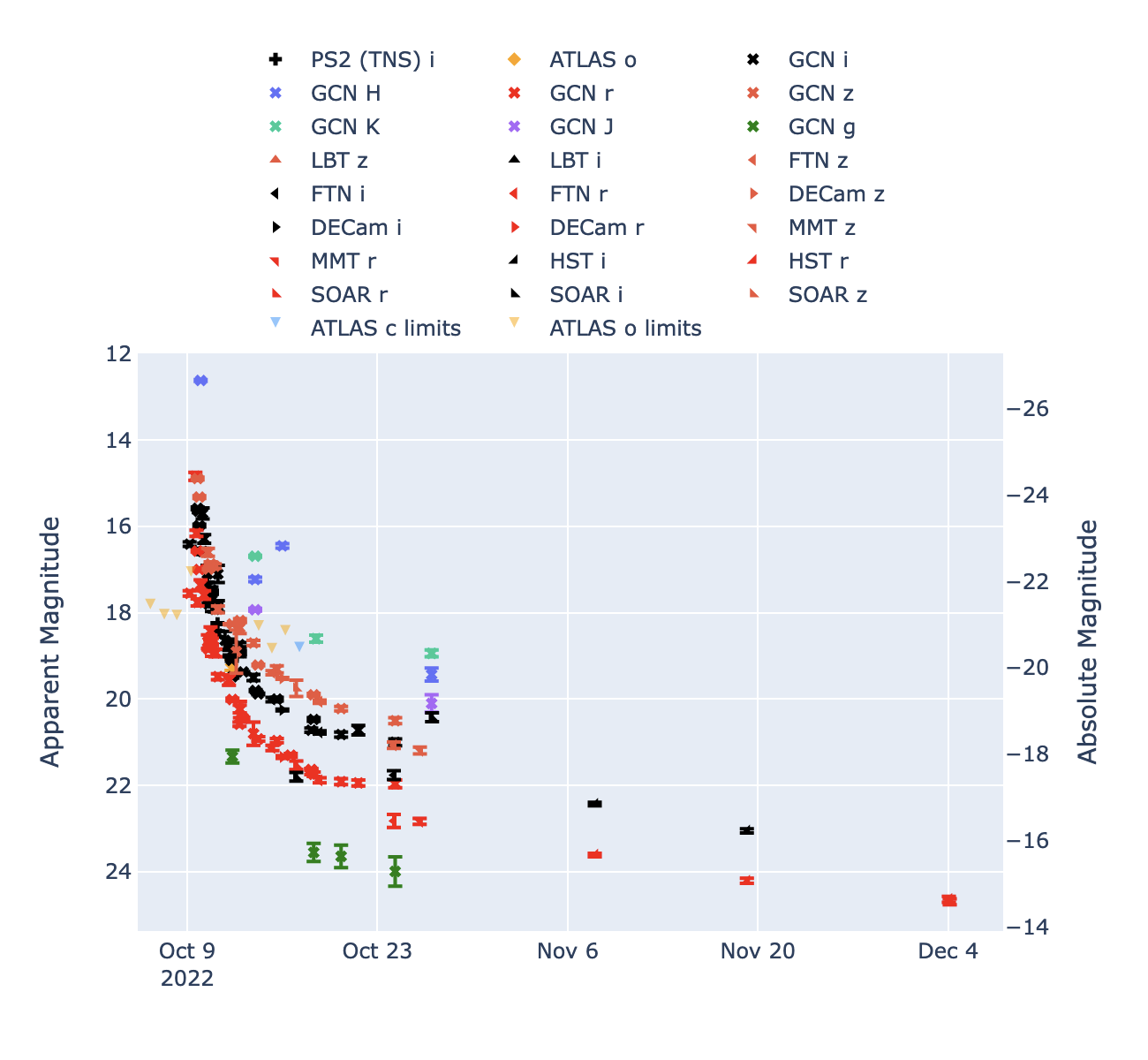}
    \caption{Light curve of GRB221009A as analyzed by \cite{shrestha_limit_2023}. These data were aggregated from GCN circulars and uploaded to the TOM in real time. An interactive version of this figure, extracted from the SAGUARO TOM, is \href{https://content.cld.iop.org/journals/0004-637X/964/1/35/revision1/apjad2170f8_int/apjad2170f8_int.html}{available}. The plot area can be zoomed and panned, and specific data series can be turned on or off by clicking in the legend.}
    \label{fig:grb}
\end{figure}

Regardless of whether they were requested through the TOM or if they were obtained by us or others, we have the ability to upload reduced data products for a given target. They will then be displayed on interactive light-curve and spectral plots on the target detail page, produced using Plotly's Python graphing library.\footnote{\url{https://plotly.com/python/}} These allow us to keep track of the evolution of targets in real time and plan or cancel follow-up accordingly. We have tested this functionality by using it to manually aggregate GRB data from GCN circulars (e.g., Figure~\ref{fig:grb}). However, we strongly encourage other observers to switch to machine-readable communications as soon as possible, so that they can be ingested into TOMs automatically (see Section~\ref{sec:hermes}). In the future, we hope to include preliminary analysis tools, such as spectral-line plotting and comparison to light-curve or spectral templates, directly in the TOM.

\subsection{Reporting Results}\label{sec:results}
\subsubsection{Treasure Map}
The Gravitational Wave Treasure Map is a tool to coordinate, visualize, and assess the EM follow-up of GW events \citep{wyatt_gravitational_2020}. We report CSS pointings within the 95\% localization region of any ``active'' GW events to the Treasure Map. We consider a GW event to be active during a given pointing if
\begin{enumerate}
    \item the event is astrophysically significant,
    \item the event has not been retracted, and
    \item the pointing occurred between 0 and 3 days after the most recent estimate of the merger time.
\end{enumerate}
Since the beginning of 2023 September, when we restarted our reporting, we have sent 287 reports of 283 serendipitous pointings for 12 BBH mergers to the Treasure Map (some observations apply to multiple events), corresponding to 1435 deg$^2$ of localization-region coverage.

To accomplish this, we have developed the \texttt{tom_treasuremap} app, which, when used alongside the \texttt{tom_nonlocalizedevents} (Section~\ref{sec:alerts}) and \texttt{tom_surveys} (Section~\ref{sec:survey}) apps, provides a \verb|TreasureMapPointing| model connecting a \verb|NonLocalizedEvent| to a \verb|SurveyObservationRecord| and labeling them with a Treasure Map pointing ID number and a status (planned or completed). When we submit observation requests to CSS, we automatically send these to the Treasure Map as planned pointings and store the pointing IDs in our database. The app also provides the management command \verb|report_pointings|, which queries the database for completed observations (requested or serendipitous) in the localization regions of any active GW events and reports them to the Treasure Map. We run this command hourly via a cronjob on the TOM server.

\subsubsection{Transient Name Server}
\begin{figure*}
    \centering
    \includegraphics[width=0.706\textwidth]{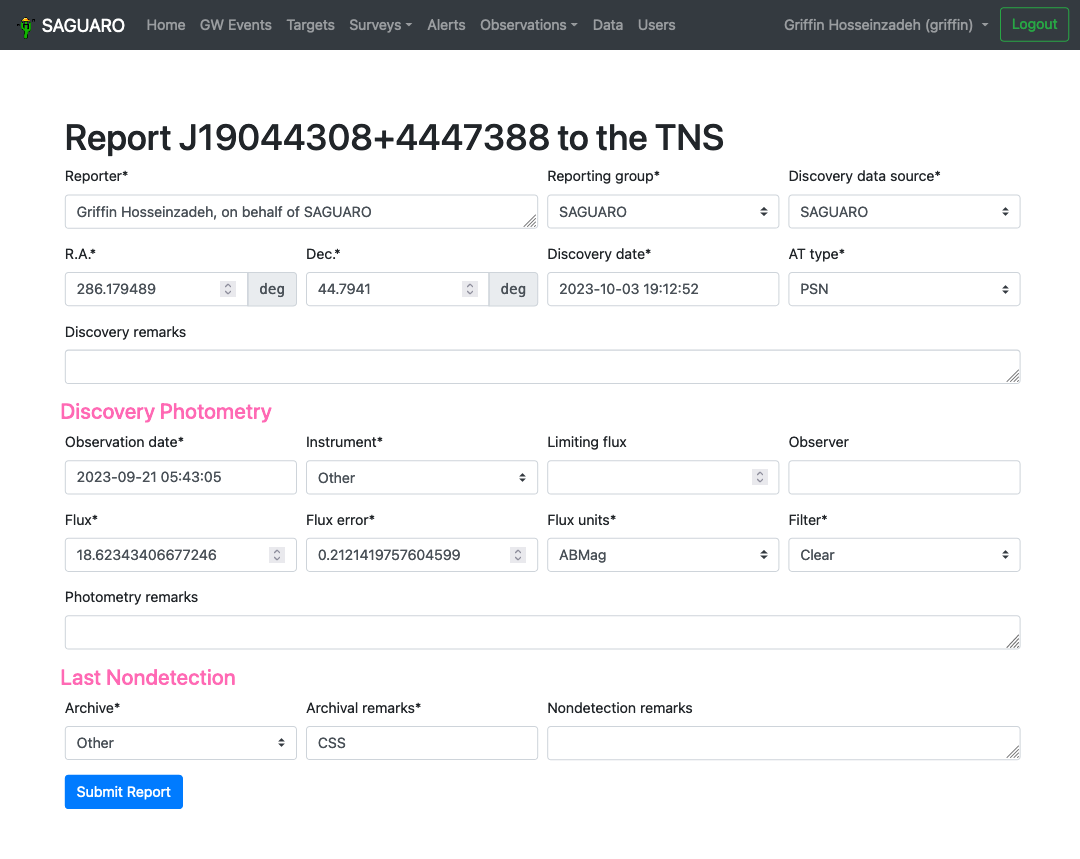}
    \includegraphics[width=0.706\textwidth]{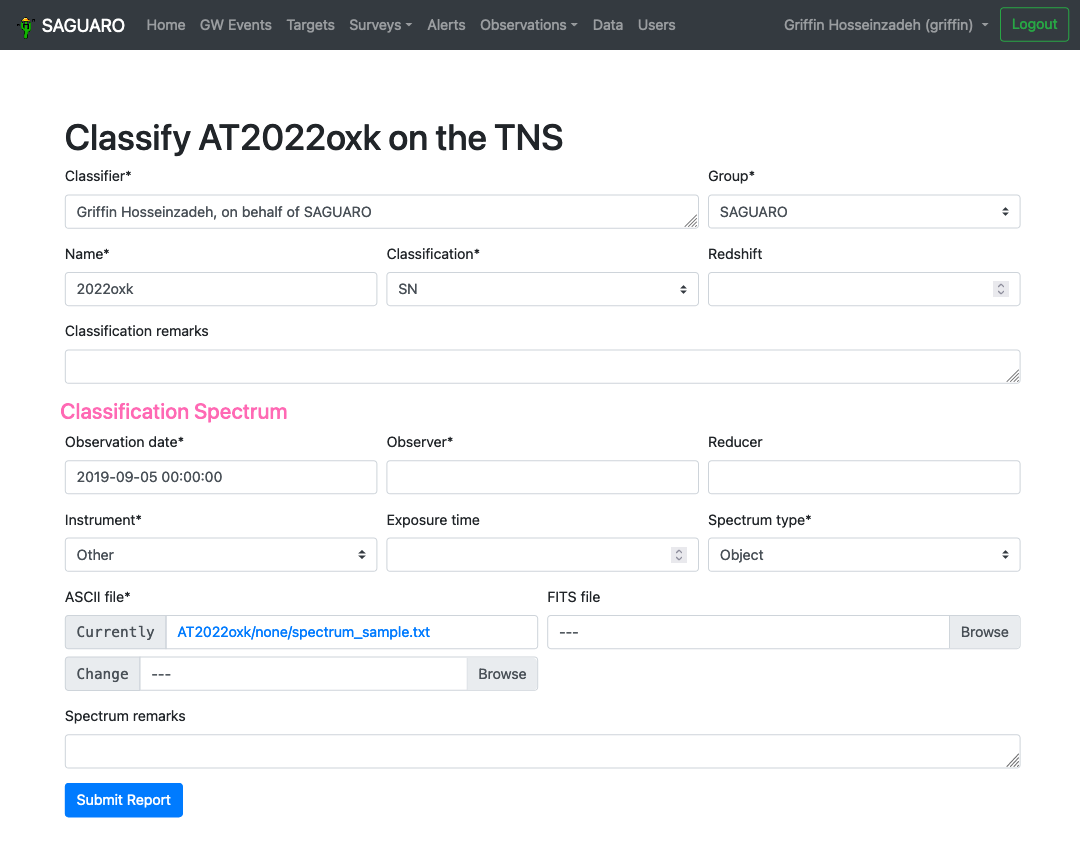}
    \caption{Forms for sending transient discovery (top) and classification (bottom) reports to the TNS from within the SAGUARO TOM. These forms are prefilled with all relevant information that is available within the TOM, including photometric and spectroscopic data. This allows the user to quickly review and submit the report with little to no typing.}
    \label{fig:tns}
\end{figure*}

The IAU Transient Name Server (TNS) is the database of record for the discovery and classification of astronomical transients \citep{gal-yam_tns_2021}. When we discover real transients in the course of our GW follow-up survey, whether or not they are real EM counterparts to GW events, we wish to report their coordinates and magnitude to the TNS and receive a permanent astronomical transient (AT) name for them. To facilitate this, we have built into our TOM a web form  (Figure~\ref{fig:tns}, top), accessed by clicking the gray ``Report'' button on the target detail page, that sends a discovery report to the TNS API without leaving the TOM. If the transient was discovered by our pipeline, the form will automatically be filled out with all the required information. Users can review the prefilled form and add authors or comments before submitting. After submission, the target will have been renamed with a permanent AT name in the TOM.

Likewise, if we obtain a classification spectrum of a transient and upload it to the TOM, we will want to report the classification to the TNS. If a target name starts with ``AT,'' a gray ``Classify'' button will replace the ``Report'' button on the target detail page (e.g., Figure~\ref{fig:target}). Clicking this button leads to a classification report form, which is also prefilled to the extent possible (Figure~\ref{fig:tns}, bottom). Submitting the form will send the report and upload its associated files to the TNS. After submission, the target's classification and redshift, if provided, will appear on the target detail page. If the target's prefix is changed (e.g., from ``AT'' to ``SN'') as a result of the classification, its name will also be updated.

\subsubsection{Hermes}\label{sec:hermes}
In addition to programmatic reporting of observations to these two databases, the GW follow-up community has traditionally relied on free-text GCN circulars to communicate information about potential EM counterparts, including their discovery, their likelihood of association with particular GW events, and the results of any vetting or follow-up observations. SCIMMA's Hermes service\footnote{\url{https://hermes.lco.global/}} is designed to supplement GCN circulars by providing several semistructured message schemata for communicating targets and data in a machine-readable format, alongside free text. Messages sent through Hermes can also be reduced to plain text and sent as GCNs at the user's request. We plan to integrate Hermes messaging into the SAGUARO TOM in a similar method to our TNS reporting---providing a form prefilled with details of targets and data from the TOM that can be quickly reviewed and sent to the Hermes API. Eventually, this will be our preferred method of composing GCN circulars, but as Hermes has been under active development during the same time period as our TOM, we have not made this a priority for early O4. Until this is integrated into the TOM, we plan to send any GCN circulars manually through the Hermes web interface.

\section{Image-subtraction Pipeline}\label{sec:pipeline}
\subsection{Deeper Reference Images}\label{sec:references}
\begin{figure}
    \centering
    \includegraphics[width=0.49\textwidth]{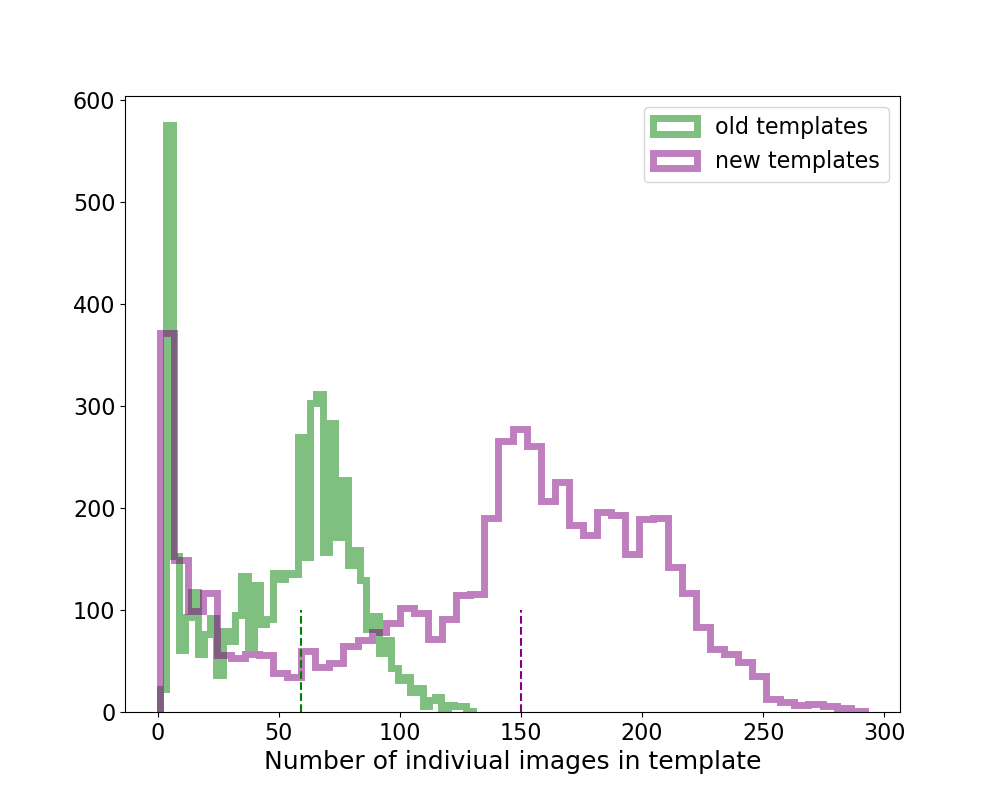}
    \caption{Histograms of the number of individual images that were used to create the reference templates for the old templates vs.\ the newly rebuilt templates. Dashed lines indicate the median values.}
    \label{fig:ref_stats}
\end{figure}

With the accumulation of more than 4~yr of data since the creation of templates for SAGUARO in 2019, we look to create new templates for image subtraction using the additional data for use during O4. Previously, we had access to a total of 5385 fields, arrayed in the sky accessible in southern Arizona, excluding areas near the Galactic plane (see Figure~1 of \citealt{lundquist_searches_2019}) with reference templates. We have new data for 4916 of these fields, allowing us to build deeper templates for 91\% of our previous fields, plus additional data for 25 new fields, expanding the number of templates. We build these new templates using the same criteria as our prior work, requiring the number of stars detected in an image to be greater than 2000, and the sky brightness lower than 20~mag (except in the cases where the total number of images is less than four, for which we implement no cuts). Compared to the previous sample of image templates that were created with a median of 60 individual images (see \citealt{lundquist_searches_2019} for details), the new sample of image templates (which includes the deeper template for fields that had additional data, the old templates for the fields that were not observed again since 2019, and the new additional fields) were created with a median of 149 images. On the low end, those with fewer than five images make up less than 1\% (compared to 9\% previously) and 74\% of templates were made with $>$90 images (compared to the previous 5\%). A histogram of the number of individual images that make up both the old and new templates is shown in Figure \ref{fig:ref_stats}. We expect these deeper reference images to result in fewer image-subtraction artifacts and allow the recovery fainter transients. We leave a full analysis of their effect for future work, after they have been in use for a longer period.

\subsection{Machine Learning}
\label{sec:ml}
Previously, SAGUARO made use of a random-forest ML algorithm \citep{breiman_random_2001}, as implemented in  \texttt{scikit-learn} \citep{pedregosa_scikit-learn_2011}, to classify candidates from the image subtraction as real (astrophysical transients, variable stars, or slow-moving objects) or bogus (artifacts caused by bad subtraction, bright stars, or detector effects). The random forest was trained on $10 \times 10$ pixels taken from the difference images centered on the source (equal to a $15\farcs4 \times 15\farcs4$ box; see \citealt{paterson_searches_2021} for details) and was able to recover 94\% of the real transients used for validation (where the recovery threshold was set to 0.7). In an attempt to improve the real--bogus score for O4, we looked at enhancing the method used for the ML. First we made use of the Zooniverse portal\footnote{\url{https://www.zooniverse.org/}} to classify good images for the transients and moving objects (through a private project). We found a sample of 383 and 2480 images for each, respectively. To increase the number of these images in our training set, we included each image rotated by 90\degr, 180\degr, and 270\degr, as well as the relative inversions through the $x$- and $y$-axes, which also ensures the algorithm is not sensitive to the orientation of the candidate. We then extracted a subset of images flagged as variable stars from our previous observations to add to the transient and moving-object sample of real detections, as well as a subset of images flagged as bad for the bogus sample in our training set.

Convolutional neural networks \citep[CNNs;][]{cnn} are often used for image classification problems and have been used extensively in astronomy for object classification, including the real--bogus classification for transients \citep[e.g.,][]{rb_ml_5,rb_ml_7,rb_ml_3,rb_ml_2,rb_ml_4,rb_ml_1,rb_ml_6}. Using TensorFlow \citep{abadi_tensorflow_2016,abadi_tensorflow_2016a}, we set up a basic CNN consisting of multiple 2D convolution layers, maxpooling, dropouts, and Dense layers (all with the relu activation), and finally ending with a Dense layer with a sigmoid activation. For the training data, we tried a combination of arrays extracted from the new, reference, difference, and corrected significance ($S_\mathrm{corr}$) images derived from the image subtraction. We tried training on the full $64 \times 64$ pixel ($99'' \times 99''$) thumbnails created from the image-subtraction pipeline, as well as on the inner $16 \times 16$ pixels ($25'' \times 25''$) centered on the source. Using the CNN, we found using the arrays just from the $S_\mathrm{corr}$ images provided the highest accuracy, while including the other arrays often resulted in a drop of accuracy due to overfitting. While the initial results were similar for the $64 \times 64$ and $16 \times 16$ pixel cutouts, further testing found that using the full $64 \times 64$ pixel array increased the false positive rate for images where bright stars were present on the edge of the cutout, while the $16 \times 16$ pixel array avoided this. We thus decided to use $16 \times 16$ pixels from the $S_\mathrm{corr}$ image as the final training data. We then looked to optimize the hyperparameters of the CNN model using this data. The final CNN model comprises two blocks of two 2D convolution layers of increasing filter size (16 for the first block and 32 for the second), followed by a $2 \times 2$ maxpooling, a 0.25 dropout, two 2D convolution layers with a filter size of 64, a reshaping filter to flatten the array, and two Dense layers with values of 64 and 32. All of these layers are applied using the relu activation. A final Dense layer with a value of 2 and a sigmoid activation is then applied, with a final dropout of 0.5. 

\begin{figure}
    \centering
    \includegraphics[width=0.75\columnwidth]{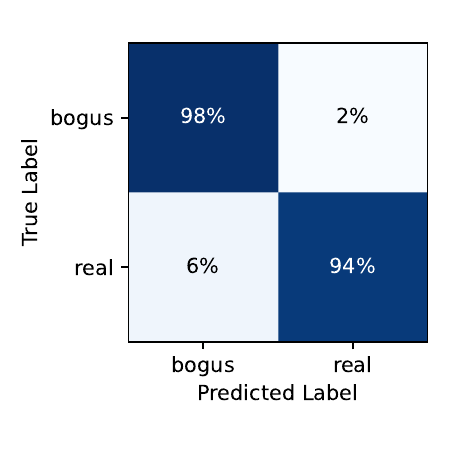}
    \caption{Confusion matrix for the new CNN ML algorithm, based on a highest probability classification; 98\% of bogus and 94\% of real detections are classified correctly.}
    \label{fig:ml_stats}
\end{figure}

The final CNN returns an array of two values containing the probability (from 0 to 1) of the candidate being real and the probability of the candidate being bogus. While the classification of real--bogus is set by whichever class has the highest probability, looking at both values can provide additional insight into candidates that the ML had trouble classifying. Using the highest probability value to classify the candidates, we correctly recover 94\% of real objects and label 98\% of the bad data as bogus (see Figure~\ref{fig:ml_stats} for the confusion matrix).

\subsection{Candidate Ingestion}\label{sec:ingestion}
Although the postprocessing procedure has not changed significantly compared to the steps described by \cite{lundquist_searches_2019} and \cite{paterson_searches_2021}, we have upgraded this portion of the pipeline to increase its speed and decrease its memory usage. We typically run the pipeline in a multithreaded mode so that we can process up to 16 images at once. In refactoring the pipeline, we have moved the reading of large files (i.e., the trained random-forest classifier and the moving-object catalog) outside of the threads, so that they are only accessed once at the beginning of the night. The trained CNN classifier is reloaded for each image, due to limitations of TensorFlow in the Python multiprocessing environment. We have also vectorized the postprocessing operations to the greatest extent possible to avoid overheads associated with Python loops. With the exception of thumbnail creation, which requires writing to the hard disk, the duration of these steps is now nearly independent of the length of the source catalog. We describe each step below.

After extracting sources from the difference images using Source Extractor \citep{bertin_sextractor:_1996}, the pipeline crossmatches the source catalog with 1.3~million solar system objects known to the Minor Planet Center. We download an up-to-date catalog of orbital elements at the beginning of each night and use PyEphem \citep{rhodes_pyephem_2011} to calculate the equatorial coordinates of each body at the time of each image. Then we use the Astropy \texttt{coordinates} module \citep{astropy_collaboration_astropy_2022} to crossmatch the source catalog with the solar system catalog, using a matching radius of $25''$ to account for uncertainties in the orbits. In total, this step takes about 50~s per image, almost all of which is spent calculating the ephemerides. Any sources with matches in the solar system catalog are marked as moving objects in our database and not displayed by default on the survey candidates page.

We then run both our old random-forest and our new CNN real--bogus classifiers, each of which takes $\lesssim$1~s per source catalog. Finally, for each source in the catalog, we produce thumbnails of the image quadruplet (see Section~\ref{sec:scanning}), which are saved on the data-reduction server and hosted with a lightweight Flask\footnote{\url{https://flask.palletsprojects.com/}} web server so that they can be displayed in the TOM. Saving the thumbnails takes a few milliseconds per source, for a total of 1--50~s per image, depending on the length of the catalog.

The pipeline records our observations in four places in the TOM database:
\begin{enumerate}
    \item the observation record table, which stores the field name, time of observation, and other metadata for each image,
    \item the targets table, which keep tracks of distinct pairs of coordinates using a $2''$ matching radius,
    \item a custom candidates table, which logs each detection of a point source along with its moving-object match and real--bogus scores, and
    \item the reduced data table, which records a photometry point for each detection.
\end{enumerate}
The new source catalog is crossmatched with the existing targets table in PostgreSQL using the Q3C extension \citep{koposov_q3c_2006} and insertions are done in bulk for each source catalog, allowing us to complete this step in a few seconds per catalog.

\begin{figure}
    \centering
    \includegraphics[width=\columnwidth]{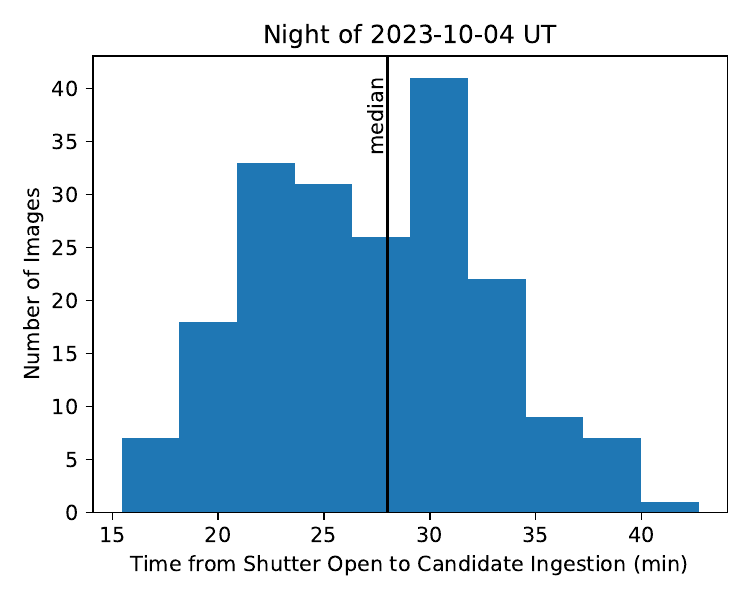}
    \caption{Histogram of the delay time between shutter open of the first of four exposures and the appearance of survey candidates in the TOM in our updated pipeline. The median delay is under 30 minutes.}
    \label{fig:delay}
\end{figure}

In total, the ingestion of candidates typically takes 1--2 minutes per image. We have also begun logging the time at which candidates are ingested into the TOM database. We find that, absent technical or network issues, the typical delay from shutter open of the first exposure to the appearance of candidates on the web page is $\lesssim$30 minutes (Figure~\ref{fig:delay}).

\section{Future Plans}
\label{sec:future}

\subsection{Bok 90Prime Upgrades}
\label{sec:90prime}
While the workhorse CSS 1.5~m telescope is a very efficient survey machine, its typical limiting magnitude ($\sim$21.5~mag) is not much deeper than the expected brightness of kilonovae ($\sim$20.5~mag) at LIGO's distance horizon for BNS mergers in O4 ($\sim$200~Mpc). Therefore, we are in the process of adding to our program a second wide-field imager on a larger telescope: 90Prime, the prime-focus imager 
on the 2.3~m (90~in) Bok Telescope on Kitt Peak, Arizona \citep{williams_90prime_2004}. 90Prime has a $\sim$1~deg$^2$ field of view and has been used for deep survey programs like the Beijing--Arizona Sky Survey \citep[BASS;][]{zou_project_2017}. Equipped with standard SDSS \textit{ugriz} filters, it also has the advantage of providing color information of transients. Over the past several years, 90Prime has undergone an upgrade, replacing its CCDs with higher-quality, blue-sensitive chips,
along with its liquid nitrogen dewar, shutter, and guider camera. Along with these hardware upgrades come upgrades to the instrument control software, as well as a new pipeline for image preprocessing and data-quality monitoring, which we discuss in Section~\ref{sec:banzai}. The upgraded 90Prime achieved engineering first light in Spring of 2023 but is now undergoing further electronics work before being released to observers. D.~Sand et al.\ (2024, in preparation) will present a complete description of the 90Prime upgrade in a future work.

\subsection{Image Processing}\label{sec:banzai}
CSS provides SAGUARO with images that have already been bias-subtracted, flat-fielded, astrometrically aligned, and photometrically calibrated. We need the equivalent preprocessing for 90Prime and any other instruments we wish to add to our survey. For this, we have adopted a customized version of Beautiful Algorithms for Normalizing Zillions of Astronomical Images \citep[BANZAI;][]{mccully_lcogt_2018}, the preprocessing pipeline developed and used by Las Cumbres Observatory. BANZAI is designed to be flexible so that it can process data from Las Cumbres' several different imagers, and our usage is a proof of concept that it can also be applied to extramural instruments with minimal modification (described below).

In addition to installing the pipeline itself, deploying BANZAI on Kitt Peak involves setting up three Docker containers \citep{merkel_docker_2014} containing a PostgreSQL database to log observations and processed data, an astrometry service based on the Astrometry.net framework \citep{lang_astrometry_2010} with index files generated from Gaia Data Release 2 \citep{gaia_collaboration_gaia_2018,lindegren_gaia_2018}, and a photometric calibration service based on the ATLAS All-sky Stellar Reference Catalog \citep{tonry_atlas_2018a}. The main customization to the pipeline for 90Prime relates to the fact that the 90Prime focal plane consists of four chips, each with four amplifiers, producing a 16-extension FITS image, whereas all Las Cumbres imagers contain a single four-amplifier chip. Our solution involves splitting each image into four four-extension FITS files and passing them to BANZAI to be processed in parallel. Relatedly, 90Prime's depth and large field of view can produce very crowded fields, which were not always a good match for the Astrometry.net configuration included with BANZAI by default. We will refine these settings further as we test the pipeline on a larger quantity of images.

\subsection{Survey Coordination}
Once the upgraded 90Prime is regularly available, we will begin to integrate it into our survey strategy. This will involve reoptimizing our field selection algorithm when 90Prime is available (usually for 1--3 weeks per lunation) to simultaneously plan observations on two telescopes. As discussed in Section~\ref{sec:survey}, we have kept the \texttt{tom_surveys} app as general as possible with this future in mind.

Likewise, we will have to adapt our image-subtraction pipeline to run on images from a new instrument, possibly using reference images from BASS. Our initial plan is to run a second instance of the pipeline on the BANZAI server on Kitt Peak to avoid the bottleneck of transferring data off the mountain. In that case, the pipeline will remotely ingest survey candidates into our TOM database and transfer only the thumbnail images off the mountain. Because of the database schema in which ``targets'' (coordinates) are tracked separately from ``candidates'' (detections; see Section~\ref{sec:ingestion}), multiple instances of the pipeline can detect the same transient, and these will automatically be associated in the TOM if they are within $2''$ of each other.

\subsection{Forced Photometry}
\cite{rastinejad_systematic_2022} and Section~\ref{sec:vetting} describe the utility of public photometry sources (e.g., ATLAS, TNS, ZTF) for vetting potential kilonovae. After several years of processing CSS data as part of SAGUARO, we ourselves have a large archive of images that we can use to rule out associations with GW events through premerger detections. We plan to set up a forced photometry service for SAGUARO--CSS images and integrate this into the TOM as part of our standard kilonova vetting procedure.

\subsection{Additional Follow-up Facilities}
In the longer term, we hope to add functionality to the TOM to trigger additional facilities accessible via API. First, we plan to add the remainder of the queue-scheduled instruments on the MMT \citep[Hectospec and Hectochelle;][]{fabricant_hectospec_2005,szentgyorgyi_hectochelle_2011} to the \texttt{tom_mmt} app. There is also ongoing work to build a TOM Toolkit facility module around the target-of-opportunity API\footnote{\url{https://www.swift.psu.edu/too_api/}} for the Neil Gehrels Swift Observatory \citep{gehrels_swift_2004}, which would be particularly useful for constraining kilonova models via very early UV photometry \citep{arcavi_first_2018}. The 4~m V\'ictor M.\ Blanco Telescope at CTIO, Chile, may also become part of AEON, allowing it to be scheduled through Las Cumbres Observatory's observing portal, providing a critical publicly accessible survey instrument in the southern hemisphere. 

\section{Summary and Outlook}
\label{sec:summary}
We have presented our completed and ongoing upgrades to SAGUARO infrastructure for optical follow-up of GW events. These include the development of a TOM that lets users receive alerts, plan and trigger a survey, scan for new survey discoveries, vet potential EM counterparts, coordinate follow-up observations, visualize reduced data, and report results. We have also described improvements to our image-subtraction pipeline, including deeper reference images, a new CNN real--bogus classifier, and streamlined candidate ingestion. Last, we described upgrades to the 90Prime wide-field imager on the 2.3~m Bok Telescope, including improvements to camera hardware, computing hardware, and software, and outlined our plans to integrate it into our search when upgrades are completed. Together, these upgrades place SAGUARO in a good position to discover and characterize EM counterparts to any BNS or NSBH mergers in the remainder of O4 and beyond, if they are visible from Arizona.

The infrastructure we have developed here has applications far beyond GW follow-up. During our design and testing, we have used the SAGUARO TOM to observe young supernovae \citep{shrestha_evidence_2024} and GRB afterglows \citep{shrestha_limit_2023}, and we plan to add astrophysical neutrino localizations \citep[e.g.,][]{icecube_collaboration_multiwavelength_2017} soon. As with the other follow-up systems we mention in Section~\ref{sec:tom}, the SAGUARO TOM will grow into a centralized portal for all transient science in our research collaboration. Likewise, we plan to extend our implementation of BANZAI to run on smaller Arizona telescopes without dedicated image processing pipelines.

We also view the SAGUARO TOM as a case study in the type of software infrastructure that will be necessary for time-domain science in the era of Vera C.\ Rubin Observatory \citep{ivezic_lsst:_2019}, when millions of transients will be discovered each year. While brokers like Alert Management, Photometry, and Evaluation of Light Curves \citep{nordin_transient_2019}, Arizona-NOIRLab Temporal Analysis and Response to Events System \citep{matheson_antares_2021}, Automatic Learning for the Rapid Classification of Events \citep{forster_automatic_2021}, Fink \citep{moller_fink_2021}, and Lasair \citep{smith_lasair_2019} will handle alert distribution, classification, and filtering, TOMs will play a critical role in managing the follow-up observations and data gathered in response. Thoughtful architecture of these systems, in coordination with other time-domain researchers, will help us maximize the scientific return from both large and small observatories.

\section*{Acknowledgments}
We thank the organizers and attendees of the GW-EM Workshop at the Schwartz/Reismann Institute for Theoretical Physics, the Gravitational Wave Physics and Astronomy Workshop 2022, and the Windows on the Universe Workshop \citep{2023_windows_on_the_universe_workshop_white_paper_working_group_windows_2024} for helpful discussions about GW alerts and coordination of EM follow-up.

Time-domain research by the University of Arizona team, M.S., and D.J.S.\ is supported by NSF grants AST-1821987, 1813466, 1908972, 2108032, and 2308181, and by the Heising-Simons Foundation under grant \#2020-1864. The Fong Group at Northwestern acknowledges support by the National Science Foundation under grant Nos.\ AST-1909358, AST-2308182, and CAREER grant No.\ AST-2047919.
K.A.B.\ is supported by an LSSTC Catalyst Fellowship; this publication was thus made possible through the support of grant 62192 from the John Templeton Foundation to LSSTC. The opinions expressed in this publication are those of the authors and do not necessarily reflect the views of LSSTC or the John Templeton Foundation. W.F.\ gratefully acknowledges support by the David and Lucile Packard Foundation, the Alfred P.\ Sloan Foundation, and the Research Corporation for Science Advancement through Cottrell Scholar Award \#28284. The TOM Toolkit is managed by Las Cumbres Observatory, with generous financial support from the Heising-Simons Foundation, the Zegar Family Foundation, and the National Science Foundation under grant No.\ 2209852.

\facilities{
ADS,
Bok,
GWTM,
SO:1.5m,
SO:Kuiper,
TNS.
}

\software{
Astrometry.net \citep{lang_astrometry_2010},
\texttt{astroplan} \citep{morris_astroplan_2018},
Astropy \citep{astropy_collaboration_astropy_2013,astropy_collaboration_astropy_2018,astropy_collaboration_astropy_2022},
\texttt{astropy-healpix} \citep{astropy_collaboration_astropy_2022},
Astroquery \citep{ginsburg_astroquery_2019},
Astro-SCRAPPY \citep{mccully_astropy_2018},
BANZAI \citep{mccully_lcogt_2018},
Beautiful Soup \citep{richardson_beautiful_2023},
\texttt{crispy-bootstrap4} \citep{smith_crispy-bootstrap4_2022},
\texttt{dateutil} \citep{niemeyer_dateutil_2021},
Django \citep{django_software_foundation_django_2023},
\texttt{django-bootstrap4} \citep{verheul_django-bootstrap4_2021},
Django ``excontrib'' Comments \citep{django_software_foundation_django_2022},
\texttt{django-crispy-forms} \citep{araujo_django-crispy-forms_2023},
Django Extensions \citep{trier_django_2023},
Django Filter \citep{gibson_django_2021},
\texttt{django-gravatar} \citep{waddington_django-gravatar_2020},
\texttt{django-guardian} \citep{balcerzak_django-guardian_2021},
\texttt{django-phonenumber-field} \citep{foulis_django-phonenumber-field_2023},
Django REST Framework \citep{christie_django_2022},
\texttt{django-webpack-loader} \citep{lone_django-webpack-loader_2022},
Docker \citep{merkel_docker_2014},
\texttt{fastavro} \citep{tebeka_fastavro_2022},
\texttt{fits2image} \citep{nation_lcogt_2022},
Flask \citep{pallets_projects_flask_2022},
Flask-SQLAlchemy \citep{pallets_projects_flask-sqlalchemy_2021},
\texttt{fundamentals} \citep{young_fundamentals_2023},
\texttt{gracedb-sdk} \citep{singer_gracedb-sdk_2022},
HEALPix Alchemy \citep{singer_healpix_2022},
\texttt{healpy} \citep{zonca_healpy_2019},
Hop Client \citep{godwin_hop_2022},
Hopskotch \citep{scimma_project_hopskotch_2023},
Light Curve Fitting \citep{hosseinzadeh_light_2023a},
\texttt{ligo.skymap} \citep{singer_rapid_2016},
\texttt{lmfit} \citep{newville_lmfit_2023},
Apache Kafka \citep{apache_software_foundation_apache_2023},
Kilonova Candidate Vetting \citep{rastinejad_kilonova_2023},
Matplotlib \citep{hunter_matplotlib:_2007},
MMT Observatory Manager \citep{gibson_queue_2018},
MOCPy \citep{baumann_cds-astro_2023},
NumPy \citep{harris_array_2020},
Paramiko \citep{forcier_paramiko_2023},
Photutils \citep{bradley_astropy_2022},
Pillow \citep{murray_python-pillow_2023},
\texttt{plotly.py} \citep{plotly_plotly_2023},
PostgreSQL \citep{postgresql_global_development_group_postgresql_2022},
Psycopg \citep{di_gregorio_psycopg_2023},
PyEphem \citep{rhodes_pyephem_2011},
pyFFTW \citep{gomersall_pyfftw_2016},
PyMMT \citep{wyatt_pymmt_2023},
Python-Markdown \citep{stienstra_python-markdown_2023},
Q3C \citep{koposov_q3c_2006},
Requests \citep{reitz_requests_2023},
SAGUARO Pipeline \citep{paterson_saguaro_2023},
SAGUARO TOM \citep{hosseinzadeh_saguaro_2023},
SASSy Q3C Models \citep{daly_sassy_2023},
SCAMP \citep{bertin_automatic_2006,scamp2},
\texttt{scikit-image} \citep{van_der_walt_scikit-image_2014},
\texttt{scikit-learn} \citep{pedregosa_scikit-learn_2011},
SciPy \citep{virtanen_scipy_2020},
\texttt{setuptools} \citep{python_packaging_authority_setuptools_2023},
\texttt{sip_tpv} \citep{shupe_more_2012},
Slack \citep{slack_technologies_slack_2023},
\texttt{slackclient} \citep{slack_technologies_slackclient_2022},
Source Extractor \citep{bertin_sextractor:_1996,bertin_sextractor_2010},
\texttt{specutils} \citep{earl_astropy_2023},
SQLAlchemy \citep{bayer_sqlalchemy_2012},
SWarp \citep{swarp},
TensorFlow \citep{abadi_tensorflow_2016a,abadi_tensorflow_2016,tensorflow_developers_tensorflow_2023},
TOM Toolkit \citep{collom_tom_2020,lindstrom_tom_2022},
\texttt{tom-alertstreams} \citep{tom_toolkit_project_tom-alertstreams_2023},
\texttt{tom_mmt} \citep{hosseinzadeh_tom_2023},
\texttt{tom_nonlocalizedevents} \citep{tom_toolkit_project_tom_nonlocalizedevents_2023},
\texttt{twilio-python} \citep{twilio_twilio-python_2023},
\texttt{urllib3} \citep{petrov_urllib3_2023},
\texttt{voevent-parse} \citep{staley_voevent-parse_2014},
Watchdog \citep{mangalapilly_watchdog_2023},
WhiteNoise \citep{evans_whitenoise_2023},
ZOGY \citep{paterson_zogy_2023}
}

\bibliography{refs,zotero_abbrev}

\end{document}

%% file: affiliation.tex
\newcommand{\NU}{\affiliation{Center for Interdisciplinary Exploration and Research in Astrophysics and Department of Physics and Astronomy, \\Northwestern University, 1800 Sherman Avenue, Floor 8, Evanston, IL 60201-3497, USA}}
\newcommand{\UA}{\affiliation{Steward Observatory, University of Arizona, 933 North Cherry Avenue, Tucson, AZ 85721-0065, USA}}
\newcommand{\Keck}{\affiliation{W.~M.~Keck Observatory, 65-1120 M\=amalahoa Highway, Kamuela, HI 96743-8431, USA}}
\newcommand{\UCDavis}{\affiliation{Department of Physics, University of California, 1 Shields Avenue, Davis, CA 95616-5270, USA}}
\newcommand{\Padova}{\affiliation{Department of Physics and Astronomy Galileo Galilei, University of Padova, Vicolo dell'Osservatorio, 3, I-35122 Padova, Italy}}
\newcommand{\INAF}{\affiliation{INAF Osservatorio Astronomico di Padova, Vicolo dell'Osservatorio 5, I-35122 Padova, Italy}}
\newcommand{\LPL}{\affiliation{Lunar and Planetary Laboratory, University of Arizona, 1629 East University Boulevard, Tucson, AZ 85721-0092, USA}}
\newcommand{\Stockholm}{\affiliation{Department of Astronomy, The Oskar Klein Center, Stockholm University, AlbaNova, 10691 Stockholm, Sweden}}
\newcommand{\UPenn}{\affiliation{Department of Physics and Astronomy, University of Pennsylvania, 209 South 33rd Street, Philadelphia, PA 19104, USA}}
\newcommand{\LCO}{\affiliation{Las Cumbres Observatory, 6740 Cortona Drive, Suite 102, Goleta, CA 93117-5575, USA}}
\newcommand{\UW}{\affiliation{Department of Astronomy, University of Washington, 3910 15th Avenue NE, Seattle, WA 98195-0002, USA}}
\newcommand{\MPA}{\affiliation{Max-Planck-Institut f\"ur Astrophysik, Karl-Schwarzschild-Stra\ss{}e 1, D-85748 Garching, Germany}}
\newcommand{\MPIA}{\affiliation{Max-Planck-Institut f\"ur Astronomie, K\"onigstuhl 17, D-69117 Heidelberg, Germany}}
\newcommand{\Catalyst}{\altaffiliation{LSSTC Catalyst Fellow}}
\newcommand{\Rubin}{\affiliation{Rubin Observatory Project Office, 950 North Cherry Avenue, Tucson, AZ 85719-4933, USA}}

%% file: authors.tex
\author[0000-0002-0832-2974]{Griffin Hosseinzadeh}
\UA

\author[0000-0001-8340-3486]{Kerry Paterson}
\MPIA

\author[0000-0002-9267-6213]{Jillian C. Rastinejad}
\NU

\author[0000-0002-4022-1874]{Manisha Shrestha}
\UA

\author{Philip N.\ Daly}
\UA

\author[0000-0001-9589-3793]{Michael J.\ Lundquist}
\Keck

\author[0000-0003-4102-380X]{David J.\ Sand}
\UA

\author[0000-0002-7374-935X]{Wen-fai Fong}
\NU

\author[0000-0002-4924-444X]{K.\ Azalee Bostroem}
\Catalyst\UA

\author[0000-0002-3841-380X]{Saarah Hall}
\NU

\author[0000-0003-2732-4956]{Samuel D.\ Wyatt}
\UW

\author[0000-0002-2575-2618]{Alex R.\ Gibbs}
\LPL

\author{Eric Christensen}
\Rubin

\author{William Lindstrom}
\LCO

\author{Jonathan Nation}
\LCO

\author[0000-0002-1278-5998]{Joseph Chatelain}
\LCO

\author[0000-0001-5807-7893]{Curtis McCully}
\LCO